\documentstyle[11pt,newpasp,twoside,epsf]{article}
\begin{document}
\title{Prospects for detecting neutral hydrogen using $21$cm radiation
from large scale structure at high redshifts}
 \author{J.S.Bagla}
\affil{Harish-Chandra Research Institute, Chhatnag Road, Jhunsi,
Allahabad 211019, India}
\author{Martin White}
\affil{Department of Physics and Astronomy, University of California,
Berkeley, CA 94720, USA}

\begin{abstract}
We estimate the signal from large scale structure at high redshifts in
redshifted $21$cm line.  We focus on $z \simeq 3$ and the $\Lambda$CDM
cosmology.  We assume that neutral hydrogen is to be found only in
galaxies, and normalise the total content to the density parameter of
neutral hydrogen in damped Lyman$\alpha$ absorption systems (DLAS).
We find 
that the {\it rms} fluctuations due to the large scale distribution of
galaxies is very small and cannot be observed at angular scales probed
by present day telescopes.  We have used the sensitivity of the Giant
meter-wave Radio Telescope (GMRT) for comparison.  We find that
observations stretching over $10^3$hours will be required for a
$3\sigma$ detection of such objects.
\end{abstract}

\section{Introduction}

Observations of galaxies and absorption systems at high redshifts have
shown that the assembly of present day galaxies started around $6 \geq
z \geq 1$.  At higher redshifts, the fraction of matter that had
collapsed to form stars was small.  By the end of the redshift range
mentioned here, the rate of star formation was starting to decline and
assembly of larger structures like the clusters of galaxies is the
dominant process.  In order to understand the process of galaxy
formation completely, it is important to study galaxies in this
redshift range in as many wavelengths as possible.

In this paper we present results for the large scale distribution of
H{\/}I at $z \simeq 3$.  More detailed results for the entire range of
redshifts will be presented elsewhere.  References for earlier
theoretical work as well as related observations can be found in Bagla
(1999a) and Bharadwaj (2001). 

Observations of galaxies and quasars at high redshifts provide
convincing proof that the inter-galactic medium is ionised out to $z
\sim 6.2$ (Pentericci et al. 2002).  Thus most of the neutral hydrogen
at $z \simeq 3$ is to be 
found in galaxies.  We have an estimate of the total neutral hydrogen
content from observations of DLAS (Storrie-Lombardi, McMahon and Irwin
, 1996).  There is no observational evidence to support the hypothesis 
that DLAS and Lyman 
break galaxies (LBG) have different properties (Fynbo et al. 2002).
Thus we can safely assume 
that neutral hydrogen is distributed uniformly amongst galaxies at $z
\simeq 3$, and that the total amount of neutral hydrogen adds up to
give us the density parameter contributed by neutral hydrogen in
DLAS. 

\begin{figure}
\plotone{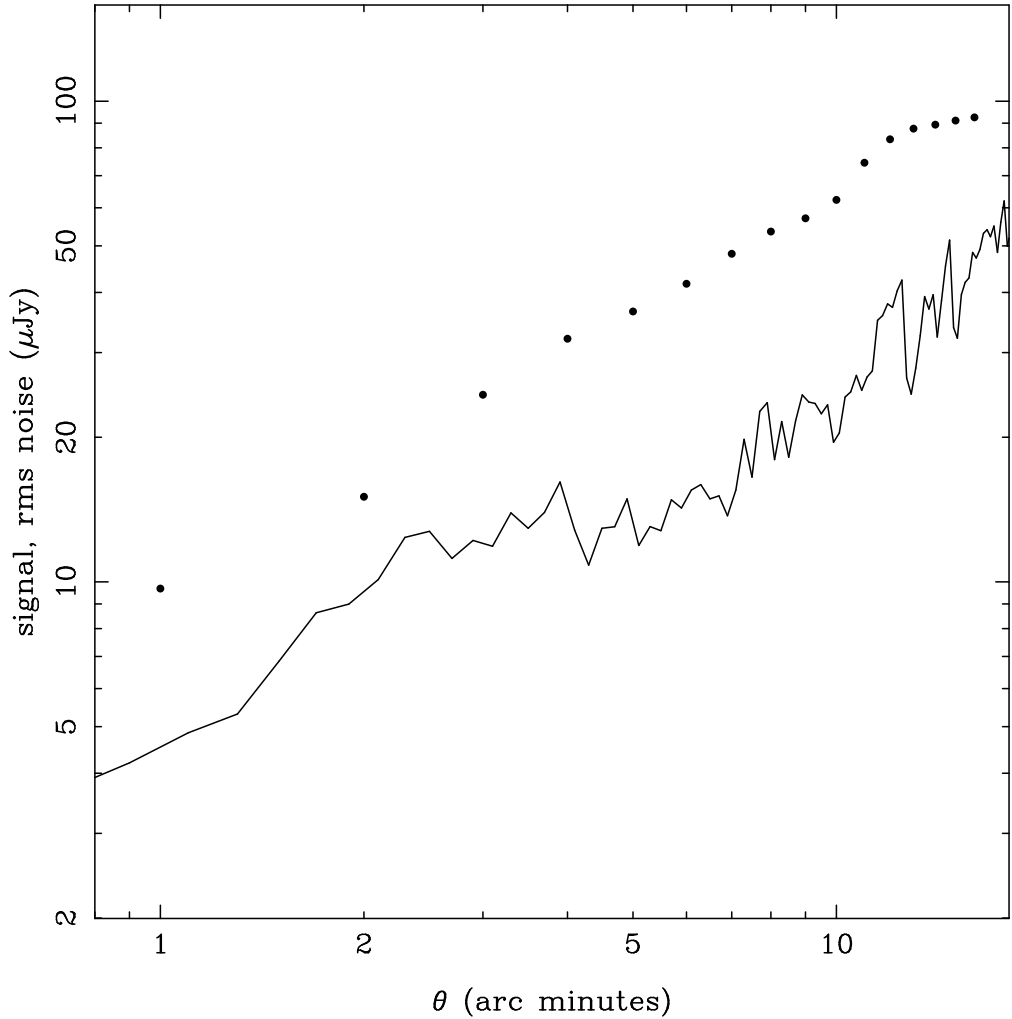}
\caption{The expected signal (dots) from largest structures at 
$z\simeq 3$ as
a function of angular scale and frequency width of $1$MHz.  Estimated
noise level (line) for the GMRT is shown as a function of scale for the 
same
bandwidth and $10^3$hours of integration.  We used the sensitivity of
GMRT at $327$MHz, which corresponds to a redshift slightly higher than
$3$.  A $3\sigma$ detection is possible at angular scales around $5'$.}
\end{figure}

\section{From Haloes to Radio Maps}

In this section we outline the method used for estimating the signal
from neutral hydrogen at high redshifts.

We ran N-Body simulations of the $\Lambda$CDM model.  The size of the
simulation box in physical units is $50h^{-1}Mpc$~(comoving) and the 
number of
particles used in this gravity only simulation was $256^3$.  We used
the TreePM method (Bagla, 1999b) for these simulations.  

The density around each particle was estimated by measuring the
distance to the $n$th neighbour, where neighbours were sorted by
distance.  The results presented here used $n=16$ but the numbers do
not change much for $n=32$ or for a different choice of estimator for
density.

Particles in regions with over-density higher than a threshold were
then selected from the simulation.  Each of these particles was
assigned an equal amount of neutral hydrogen such that the density
parameter of neutral hydrogen was $0.002$.  Results can be scaled
trivially if one prefers a different value of $\Omega_{H{\/}I}$.

The dependence of emission on the local spin temperature is weak enough to
be ignored in physical situations of interest.  The optical depth in
$21$cm is sufficiently small for us to ignore any absorption.  Thus
the problem of making radio maps essentially reduces to that of
assigning frequency and angular position to each particle in high
density regions and adding up the signal in relevant frequency channel
and pixel.  The conversion from neutral hydrogen mass to signal is
\begin{equation}
S_\nu = 309 \mu Jy \left({M_{HI} \over 10^{13} M_\odot}\right)
\left({1 {\rm MHz} \over \Delta\nu}\right).
\end{equation}

The simulated radio maps can then be used to look for optimum
frequency window and angular scales at which a search for signal
should be carried out.  In such an exercise, parameters of present
day telescopes need to be considered and we have used sensitivity
levels of the GMRT (Swarup et al. 1991) for this.  Figure~1 shows the
expected signal from 
the largest objects for a frequency channel of $1$MHz as a function of
angular scale.  We scanned the radio maps with a very fine resolution
in order to locate the maximum signal.  The expected signal is
compared with the sensitivity of the GMRT at $327$MHz for a $10^3$hour
observation.  This figure suggests that a $3\sigma$ detection of the
largest structures is possible at angular scales $3'-6'$ in such an
observation.  Signal expected from typical structures is much
smaller. 

The expected observation time is very large and hence it is important
to ask whether such extreme structures are likely to be there in the
volume sampled by the GMRT beam or not.  The volume sampled by a GMRT
beam at $327$MHz is much larger than the volume of the simulation used
here.  It is also much larger than the volume of the fields in which
spikes in the redshift distribution of LBGs have been
observed (Steidel et al. 1998).  As the rate at which spikes occur in
the redshift 
distribution of LBGs is close to one per field, we expect that the
GMRT beam will contain at least one extreme object. 

\section{Discussion}

We have used N-Body simulations to estimate the signal in redshifted
$21$cm line from large scale structure at $z \simeq 3$.  We find that
the present day telescopes can detect extreme objects in the large
scale structure at these redshifts.

In our estimation, we made use of gravity only simulations and ignored
any effects coming from gas physics.  These effects play a very
important role in the distribution and state of gases at small
scales.  However, we are interested only in gross properties and that
too at large scales and there is no reason to suspect that we will get
these wrong in the method that we have used as long as we restrain our
estimates to scales larger than $1$Mpc (comoving).  The weakest point
in our method is that we assumed that the amount of neutral hydrogen
assigned to an N-Body particle did not depend on the mass of the
collapsed structure that contained this particle.  We expect larger
structures ($M > 10^{13}$M$_\odot$) to be like groups of galaxies and
have less neutral hydrogen by fraction.  Also, we expect smaller
structures ($M < 10^{10}$M$_\odot$) to have less neutral fraction as
these will be influenced strongly by the photo-ionising background.
However, the fraction of mass in haloes at these extremes is
negligible and hence we do not expect these effects to invalidate our
results at these redshifts.  Such effects will be very important at
lower redshifts where very massive haloes are more common, or at high
redshifts where the low mass haloes contain most of the mass in
collapsed structures.

\acknowledgements

JSB thanks Jayaram Chengalur for many useful discussions.
A part of the work presented here was done using the Beowulf cluster
at the Harish-Chandra Research Institute ({\sl
http://cluster.mri.ernet.in}\/).

\end{document}